\documentclass[a4paper]{jpconf} 
\usepackage{graphicx}
\begin{document}

\title{Properties of the partonic phase at RHIC \\ within PHSD}

\author{E. L. Bratkovskaya} 
\address{Institute for Theoretical Physics, University of Frankfurt, Frankfurt, Germany}
\address{Frankfurt Institute for Advanced Study, Frankfurt, Germany}
\ead{Elena.Bratkovskaya@th.physik.uni-frankfurt.de}

\author{W. Cassing}
\address{Institute for Theoretical Physics, University of Giessen, Giessen, Germany}

\author{V. P. Konchakovski}
\address{Institute for Theoretical Physics, University of Giessen, Giessen, Germany}
\address{Bogolyubov Institute for Theoretical Physics, Kiev, Ukraine}

\author{O. Linnyk}
\address{Institute for Theoretical Physics, University of Giessen, Giessen, Germany}

\author{V. Ozvenchuk}
\address{Frankfurt Institute for Advanced Study, Frankfurt, Germany}

\author{V. Voronyuk}
\address{Joint Institute for Nuclear Research, Dubna, Russia}
\address{Frankfurt Institute for Advanced Study, Frankfurt, Germany}
\address{Bogolyubov Institute for Theoretical Physics, Kiev, Ukraine}

\begin{abstract}
The dynamics of partons, hadrons and strings in relativistic
nucleus-nucleus collisions is analyzed within the novel
Parton-Hadron-String Dynamics (PHSD) transport approach, which is
based on a dynamical quasiparticle model for partons (DQPM) matched to
reproduce recent lattice-QCD results -- including the partonic equation
of state -- in thermodynamic equilibrium. The transition from partonic
to hadronic degrees of freedom is described by covariant transition
rates for the fusion of quark-antiquark pairs or three quarks
(antiquarks), respectively, obeying flavor current-conservation, color
neutrality as well as energy-momentum conservation.  In order to
explore the space-time regions of 'partonic matter' the PHSD approach
is applied to nucleus-nucleus collisions from SPS to RHIC
energies. Detailed comparisons are presented for hadronic rapidity
spectra and transverse mass distributions. The traces of partonic
interactions are found in particular in the elliptic flow of hadrons
as well as in an approximate quark-number scaling at the top RHIC
energy.
\end{abstract}

\section{Introduction}

The 'Big Bang' scenario implies that in the first micro-seconds of the
universe the entire state has emerged from a partonic system of
quarks, antiquarks and gluons -- a quark-gluon plasma (QGP) -- to
color neutral hadronic matter consisting of interacting hadronic
states (and resonances) in which the partonic degrees of freedom are
confined. The nature of confinement and the dynamics of this phase
transition has motivated a large community for several decades and is
still an outstanding question of todays physics. Early concepts of the
QGP were guided by the idea of a weakly interacting system of partons
which might be described by perturbative QCD (pQCD). However,
experimental observations at the Relativistic Heavy Ion Collider
(RHIC) indicated that the new medium created in ultrarelativistic
Au+Au collisions is interacting more strongly than hadronic matter and
consequently this concept had to be severely questioned. Moreover, in
line with theoretical studies in Refs.~\cite{Shuryak,Thoma,Andre} the
medium showed phenomena of an almost perfect liquid of
partons~\cite{STARS,Miklos3} as extracted from the strong radial
expansion and the scaling of elliptic flow $v_2(p_T)$ of mesons and
baryons with the number of constituent quarks and
antiquarks~\cite{STARS}.

The question about the properties of this (nonperturbative) QGP liquid
is discussed controversially in the literature and dynamical concepts
describing the formation of color neutral hadrons from colored partons
are scarce. A fundamental issue for hadronization models is the
conservation of 4-momentum as well as the entropy problem because by
fusion/coalescence of massless (or low constituent mass) partons to
color neutral bound states of low invariant mass (e.g. pions) the
number of degrees of freedom and thus the total entropy is reduced in
the hadronization process~\cite{Koal1,Koal2,AMPT}. This problem - a
violation of the second law of thermodynamics as well as the
conservation of four-momentum and flavor currents - has been addressed
in Ref.~\cite{PRC08} on the basis of the DQPM employing covariant
transition rates for the fusion of 'massive' quarks and antiquarks to
color neutral hadronic resonances or strings. In fact, the dynamical
studies for an expanding partonic fireball in Ref.~\cite{PRC08}
suggest that the latter problems have come to a practical solution.

A consistent dynamical approach - valid also for strongly interacting
systems - can be formulated on the basis of Kadanoff-Baym (KB)
equations~\cite{Sascha1} or off-shell transport equations in
phase-space representation,
respectively~\cite{Sascha1,Juchem,Knoll1}. In the KB theory the field
quanta are described in terms of dressed propagators with complex
selfenergies.  Whereas the real part of the selfenergies can be
related to mean-field potentials (of Lorentz scalar, vector or tensor
type), the imaginary parts provide information about the lifetime
and/or reaction rates of time-like 'particles'~\cite{Crev}. Once the
proper (complex) selfenergies of the degrees of freedom are known the
time evolution of the system is fully governed by off-shell transport
equations (as described in Refs.~\cite{Sascha1,Crev}).  The
determination/extraction of complex selfenergies for the partonic
degrees of freedom has been performed before in
Refs.~\cite{Cassing06,Cassing07} by fitting lattice QCD (lQCD) 'data'
within the Dynamical QuasiParticle Model (DQPM). In fact, the DQPM
allows for a simple and transparent interpretation of lattice QCD
results for thermodynamic quantities as well as correlators and leads
to effective strongly interacting partonic quasiparticles with broad
spectral functions.  For a review on off-shell transport theory and
results from the DQPM in comparison to lQCD we refer the reader to
Ref.~\cite{Crev}.

The actual implementations in the PHSD transport approach have been
presented in detail in Refs.~\cite{CaBra09,PHSD-RHIC}. Here we report
again on the actual description of hadronization
(Section~\ref{Sec:hadronization}). We also present results for
rapidity distributions, transverse mass spectra and elliptic flow for
heavy ion collisions at SPS (Section~\ref{Sec:SPS}) and RHIC energies
(Section~\ref{Sec:RHIC}) in comparison to data from the experimental
collaborations.

\section{Hadronization in PHSD}
\label{Sec:hadronization}

The hadronization, i.e. the transition from partonic to hadronic
degrees of freedom, is described in PHSD by local covariant transition
rates as introduced in Ref.~\cite{PRC08} e.g. for $q+\bar{q}$ fusion
to a meson $m$ of four-momentum $p= (\omega, {\bf p})$ at space-time
point $x=(t,{\bf x})$:
\begin{eqnarray}
&&\phantom{a}\hspace*{-5mm} \frac{d N_m(x,p)}{d^4x d^4p}= Tr_q
Tr_{\bar q} \
  \delta^4(p-p_q-p_{\bar q}) \
  \delta^4\left(\frac{x_q+x_{\bar q}}{2}-x\right) \nonumber\\
&& \times \omega_q \ \rho_{q}(p_q)
   \  \omega_{\bar q} \ \rho_{{\bar q}}(p_{\bar q})
   \ |v_{q\bar{q}}|^2 \ W_m(x_q-x_{\bar q},(p_q-p_{\bar q})/2) \nonumber \\
&& \times N_q(x_q, p_q) \
  N_{\bar q}(x_{\bar q},p_{\bar q}) \ \delta({\rm flavor},\, {\rm color}).
\label{trans}
\end{eqnarray}
In Eq.~(\ref{trans}) we have introduced the shorthand notation,
\begin{equation}
Tr_j = \sum_j \int d^4x_j \int \frac{d^4p_j}{(2\pi)^4} \ ,
\end{equation}
where $\sum_j$ denotes a summation over discrete quantum numbers
(spin, flavor, color); $N_j(x,p)$ is the phase-space density of
parton $j$ at space-time position $x$ and four-momentum $p$.  In
Eq.~(\ref{trans}) $\delta({\rm flavor},\, {\rm color})$ stands
symbolically for the conservation of flavor quantum numbers as
well as color neutrality of the formed hadron $m$ which can be
viewed as a color-dipole or 'pre-hadron'.  Furthermore, $v_{q{\bar
q}}(\rho_p)$ is the effective quark-antiquark interaction  from
the DQPM  (displayed in Fig. 10 of Ref.~\cite{Cassing07}) as a
function of the local parton ($q + \bar{q} +g$) density $\rho_p$
(or energy density). Furthermore, $W_m(x,p)$ is the dimensionless phase-space
distribution of the formed 'pre-hadron', i.e.
\begin{equation} \label{Dover} W_m(\xi,p_\xi) =
\exp\left( \frac{\xi^2}{2 b^2} \right)\ \exp\left( 2 b^2 (p_\xi^2- (M_q-M_{\bar
q})^2/4) \right)
\end{equation} 
with $\xi = x_1-x_2 = x_q - x_{\bar q}$ and $p_\xi = (p_1-p_2)/2 =
(p_q - p_{\bar q})/2$. The width parameter $b$ is fixed by
$\sqrt{\langle r^2 \rangle} = b$ = 0.66 fm (in the rest frame) which
corresponds to an average rms radius of mesons. We note that the
expression~(\ref{Dover}) corresponds to the limit of independent
harmonic oscillator states and that the final hadron-formation rates
are approximately independent of the parameter $b$ within reasonable
variations. By construction the quantity~(\ref{Dover}) is Lorentz
invariant; in the limit of instantaneous 'hadron formation',
i.e. $\xi^0=0$, it provides a Gaussian dropping in the relative
distance squared $({\bf r}_1 - {\bf r}_2)^2$. The four-momentum
dependence reads explicitly (except for a factor $1/2$)
\begin{equation} 
(E_1 - E_2)^2 - ({\bf p}_1 - {\bf p}_2)^2 -
(M_1-M_2)^2 \leq 0
\end{equation} 
and leads to a negative argument of the second
exponential in (\ref{Dover}) favoring the fusion of partons with
low relative momenta $p_q - p_{\bar q}= p_1-p_2$.

Related transition rates (to Eq.~(\ref{trans})) are defined for
the fusion of three off-shell quarks ($q_1+q_2+q_3 \leftrightarrow
B$) to color neutral baryonic ($B$ or $\bar{B}$) resonances of
finite width (or strings) fulfilling energy and momentum
conservation as well as flavor current conservation using Jacobi coordinates
(cf. Ref.~\cite{CaBra09}).

On the hadronic side the PHSD transport approach includes explicitly
the baryon octet and decouplet, the $0^-$- and $1^-$-meson nonets as
well as selected higher resonances as in HSD~\cite{HSD}. Hadrons of
higher masses ($>$ 1.5 GeV in case of baryons and $>$ 1.3 GeV in case
of mesons) are treated as 'strings' (color-dipoles) that decay to the
known (low-mass) hadrons according to the JETSET
algorithm~\cite{JETSET}. We discard an explicit recapitulation of the
string decay and refer the reader to the original work~\cite{JETSET}
or Ref.~\cite{Falter}.

\section{Application to nucleus-nucleus collisions at SPS energies}
\label{Sec:SPS}

\begin{figure}[tbh]
  \includegraphics[width=0.7\textwidth]{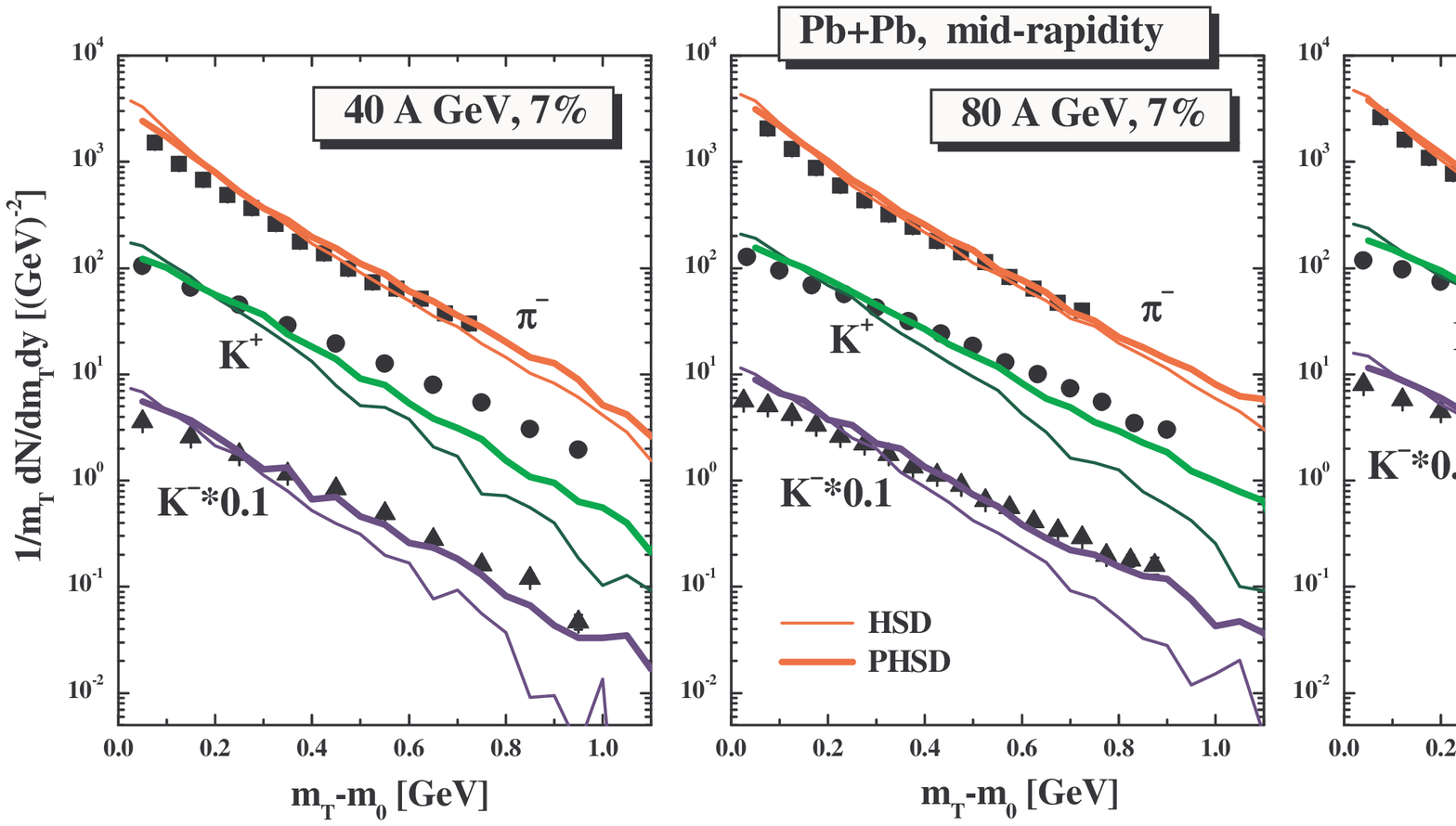}
  \caption{The $\pi^-$, $K^+$ and $K^-$ transverse mass spectra for
    central Pb+Pb collisions at 40, 80 and 158 A$\cdot$GeV from PHSD
    (thick solid lines) in comparison to the distributions from HSD
    (thin solid lines) and the experimental data from the NA49
    Collaboration~\cite{NA49a}. }
  \label{fig11}
\end{figure}

In this Section we employ the PHSD approach to nucleus-nucleus
collisions at moderate relativistic energies.  It is of interest, how
the PHSD approach compares to the HSD~\cite{HSD} model (without
explicit partonic degrees-of-freedom) as well as to experimental
data. In Fig.~\ref{fig11} we show the transverse mass spectra of
$\pi^-$, $K^+$ and $K^-$ mesons for 7\% central Pb+Pb collisions at 40
and 80 A$\cdot$GeV and 5\% central collisions at 158 A$\cdot$GeV in
comparison to the data of the NA49 Collaboration~\cite{NA49a}.  Here
the slope of the $\pi^-$ spectra is only slightly enhanced in PHSD
relative to HSD which demonstrates that the pion transverse motion
shows no sizeable sensitivity to the partonic phase. However, the
$K^\pm$ transverse mass spectra are substantially hardened with
respect to the HSD calculations at all bombarding energies - i.e. PHSD
is more in line with the data - and thus suggests that partonic
effects are better visible in the strangeness-degrees of freedom. The
hardening of the kaon spectra can be traced back to parton-parton
scattering as well as a larger collective acceleration of the partons
in the transverse direction due to the presence of repulsive vector
fields for the partons. The enhancement of the spectral slope for
kaons and antikaons in PHSD due to collective partonic flow shows up
much clearer for the kaons due to their significantly larger mass
(relative to pions). We recall that in Refs.~\cite{BratPRL} the
underestimation of the $K^\pm$ slope by HSD (and also UrQMD) had been
suggested to be a signature for missing partonic degrees of freedom;
the present PHSD calculations support this early suggestion. Moreover,
the PHSD calculations for RHIC energies show a very similar trend -
the inverse slope increases by including the partonic phase.

The strange antibaryon sector is of further interest since here the HSD
calculations have always underestimated the yield~\cite{Geiss}.  Our
detailed studies in Ref.~\cite{CaBra09} show that the HSD and PHSD
calculations both give a reasonable description of the $\Lambda +
\Sigma^0$ yield of the NA49 Collaboration~\cite{NA49_aL09}; both models
underestimate the NA57 data~\cite{NA57} by about 30\%. An even larger
discrepancy in the data from the NA49 and NA57 Collaborations is seen
for $(\bar \Lambda + \bar \Sigma^0)/N_{wound}$; here the PHSD
calculations give results which are in between the NA49 data and the
NA57 data whereas  HSD underestimates the $(\bar \Lambda + \bar
\Sigma^0)$ midrapidity yield at all centralities.

The latter result suggests that the partonic phase does not show up
explicitly in an enhanced production of strangeness (or in particular
strange mesons and baryons) but leads to a different redistribution of
antistrange quarks between mesons and antibaryons.  In fact, as
demonstrated in Ref.~\cite{CaBra09}, we find no sizeable differences
in the double strange baryons from HSD and PHSD -- in a good agreement
with the NA49 data -- but observe a large enhancement in the double
strange antibaryons for PHSD relative to HSD.

\section{Application to nucleus-nucleus collisions at RHIC energies}
\label{Sec:RHIC}

\begin{figure}[tbh]
\centerline{\includegraphics[width=0.9\textwidth]{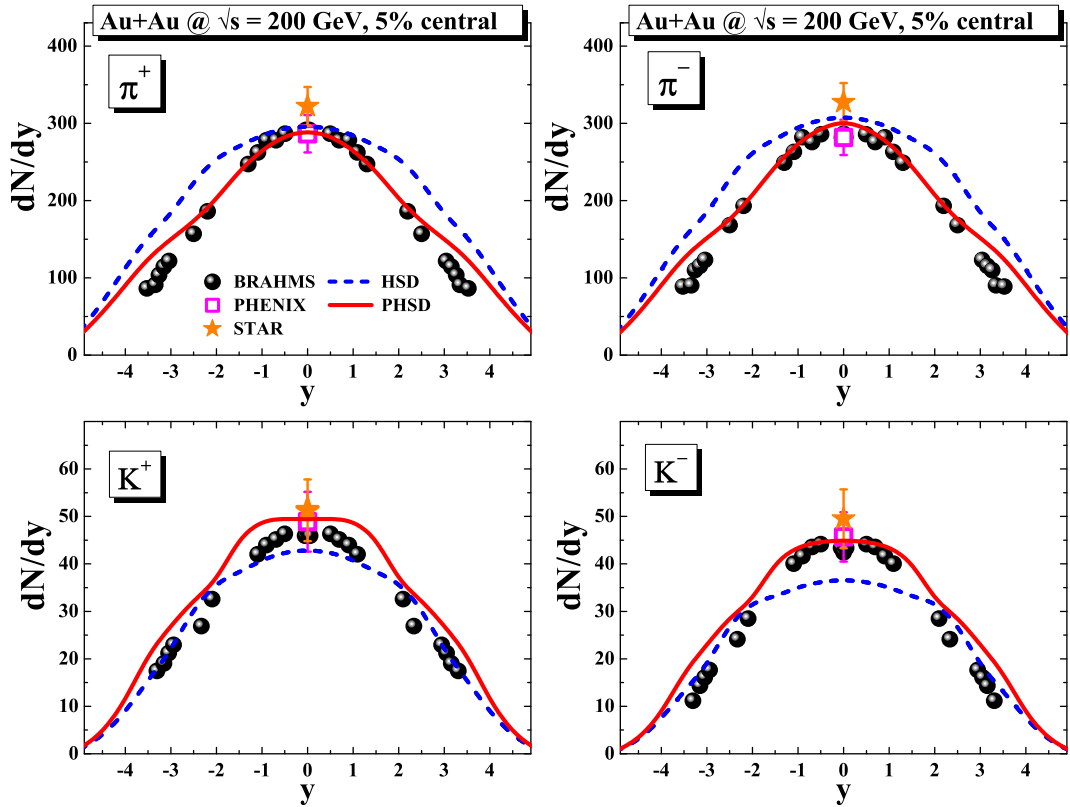}}
\caption{The rapidity distribution of $\pi^+$ (upper part, l.h.s.),
  $K^+$ (lower part, l.h.s.), $\pi^-$ (upper part, r.h.s.) and $K^-$
  (lower part, r.h.s.) for 5\% central Au+Au collisions at $\sqrt{s}$
  = 200 GeV from PHSD (solid red lines) in comparison to the
  distribution from HSD (dashed blue lines) and the experimental data
  from the RHIC Collaborations~\cite{PHENIX2,STAR3,BRAHMS}.}
\label{fig13}
\end{figure}

In this section we continue the comparison of the PHSD transport
approach to the experimental data from the RHIC collaborations as well
as to the correspondent HSD results~\cite{PHSD-RHIC}.
We find the rapidity distributions of the charged mesons to be
slightly narrower in PHSD than those from HSD and actually closer to
the experimental data. Also note that there is slightly more
production of $K^\pm$ mesons in PHSD than in HSD while the number of
charged pions is slightly lower. The actual deviations between the
PHSD and HSD spectra are not dramatic but more clearly visible than at
SPS energies (cf. Ref.~\cite{CaBra09}).  Nevertheless, it becomes
clear from Fig.~\ref{fig13} that the energy transfer in the
nucleus-nucleus collision from initial nucleons to produced hadrons --
reflected dominantly in the light meson spectra -- is rather
accurately described by PHSD. Fig.~\ref{fig13} also demonstrates that
the longitudinal motion is well understood within the PHSD approach.

\begin{figure}[tbh]
\centerline{\includegraphics[width=0.6\textwidth]{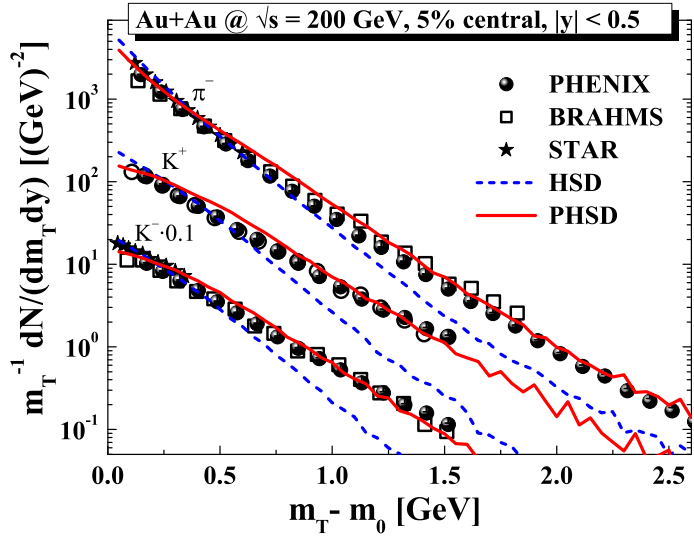}}
\caption{The $\pi^-$, $K^+$ and $K^-$ transverse mass spectra for 5\%
  central Au+Au collisions at $\sqrt{s}$ = 200 GeV from PHSD (solid
  red lines) in comparison to the distributions from HSD (dashed blue
  lines) and the experimental data from the BRAHMS, PHENIX and STAR
  Collaborations~\cite{PHENIX2,STAR3,BRAHMS} at midrapidity.}
\label{fig14}
\end{figure}

Independent information on the active degrees of freedom is provided
by transverse mass spectra of the hadrons especially in central
collisions.  The actual results for RHIC energies are displayed in
Fig.~\ref{fig14} where we show the transverse mass spectra of $\pi^-$,
$K^+$ and $K^-$ mesons for 5\% central Au+Au collisions at $\sqrt{s}$
= 200 GeV in comparison to the data of the RHIC
Collaborations~\cite{PHENIX2,STAR3,BRAHMS}.  Here the slope of the
$\pi^-$ spectra is slightly enhanced in PHSD (solid red lines)
relative to HSD (dashed blue lines) which demonstrates that the pion
transverse mass spectra also show some sensitivity to the partonic
phase (contrary to the SPS energy regime). The $K^\pm$ transverse mass
spectra are substantially hardened with respect to the HSD
calculations -- i.e. PHSD is more in line with the data -- and thus
suggest that partonic effects are better visible in the strangeness
degrees-of-freedom. The hardening of the kaon spectra can be traced
back to parton-parton scattering as well as a larger collective
acceleration of the partons in the transverse direction due to the
presence of the repulsive mean-field for the partons.
The enhancement of the spectral slopes for kaons and antikaons in PHSD
due to collective partonic flow shows up much clearer for the kaons
due to their significantly larger mass (relative to pions).

Of additional interest are the collective properties of the strongly
interacting system which are explored experimentally via the elliptic
flow
\begin{equation} 
v_2(p_T,y) = \left\langle (p_x^2 - p_y^2)/(p_x^2 +
p_y^2) \right\rangle |_{p_T,y}
\end{equation} 
of hadrons as a function of centrality, rapidity $y$, transverse
momentum $p_T$ or transverse kinetic energy per participating quarks
and antiquarks. We note that the reaction plane in PHSD is given by
the $x-z$ plane with the $z$-axis in beam direction.

\begin{figure}[tbh]
\centerline{ \includegraphics[width=0.6\textwidth]{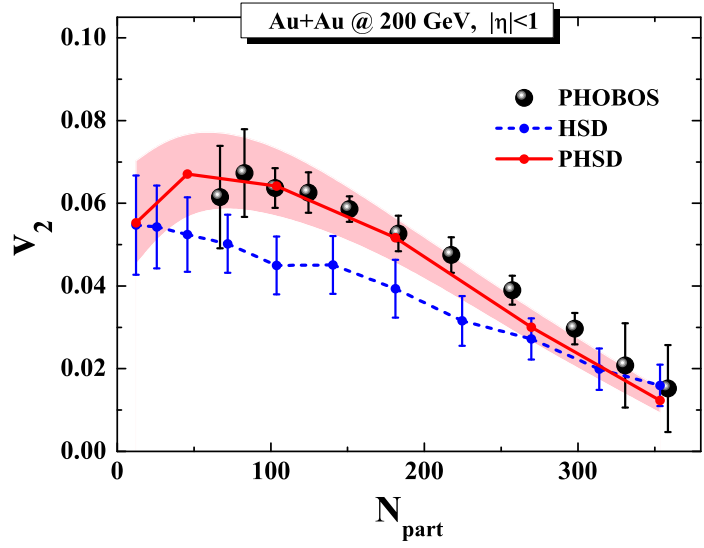}}
\caption{ The elliptic flow $v_2$ for Au+Au collisions at the top RHIC
  energy $\sqrt{s}$ = 200 GeV as a function of the centrality measured
  by the number of participating nucleons $N_{part}$. The solid (red)
  line stands for the results from PHSD whereas the dashed (blue) line
  represents the results from HSD (from Ref. \cite{Brat03}).  The data
  are taken from the PHOBOS Collaboration \cite{PHOBOS4} and
  correspond to momentum integrated events in the pseudo-rapidity
  window $|\eta| \leq 1$ for charged particles. The shaded band
  signals the statistical uncertainties of the PHSD calculations.}
 \label{fig22}
\end{figure}

We start in Fig.~\ref{fig22} with the elliptic flow $v_2$ (for Au+Au
collisions at the top RHIC energy) as a function of the centrality of
the reaction measured by the number of participating nucleons
$N_{part}$. The solid (red) line stands for the results from PHSD
which is compared to the data for charged particles from the PHOBOS
Collaboration~\cite{PHOBOS4}. The dashed blue line refers to the
corresponding results for $v_2$ from HSD (taken from
Ref.~\cite{Brat03}). The momentum integrated results in the
pseudo-rapidity window $|\eta| \leq 1$ from PHSD compare well to the
data from Ref.~\cite{PHOBOS4} whereas the HSD results clearly
underestimate the elliptic flow as pointed out before in
Ref.~\cite{Brat03}. The relative enhancement of $v_2$ in PHSD with
respect to HSD can be traced back to the high interaction rate in the
partonic phase and to the repulsive mean field for partons. We note in
passing that PHSD calculations without mean fields only give a tiny
enhancement for the elliptic flow relative to HSD.

\begin{figure}[t]
  \centerline{
    \includegraphics[width=0.5\textwidth]{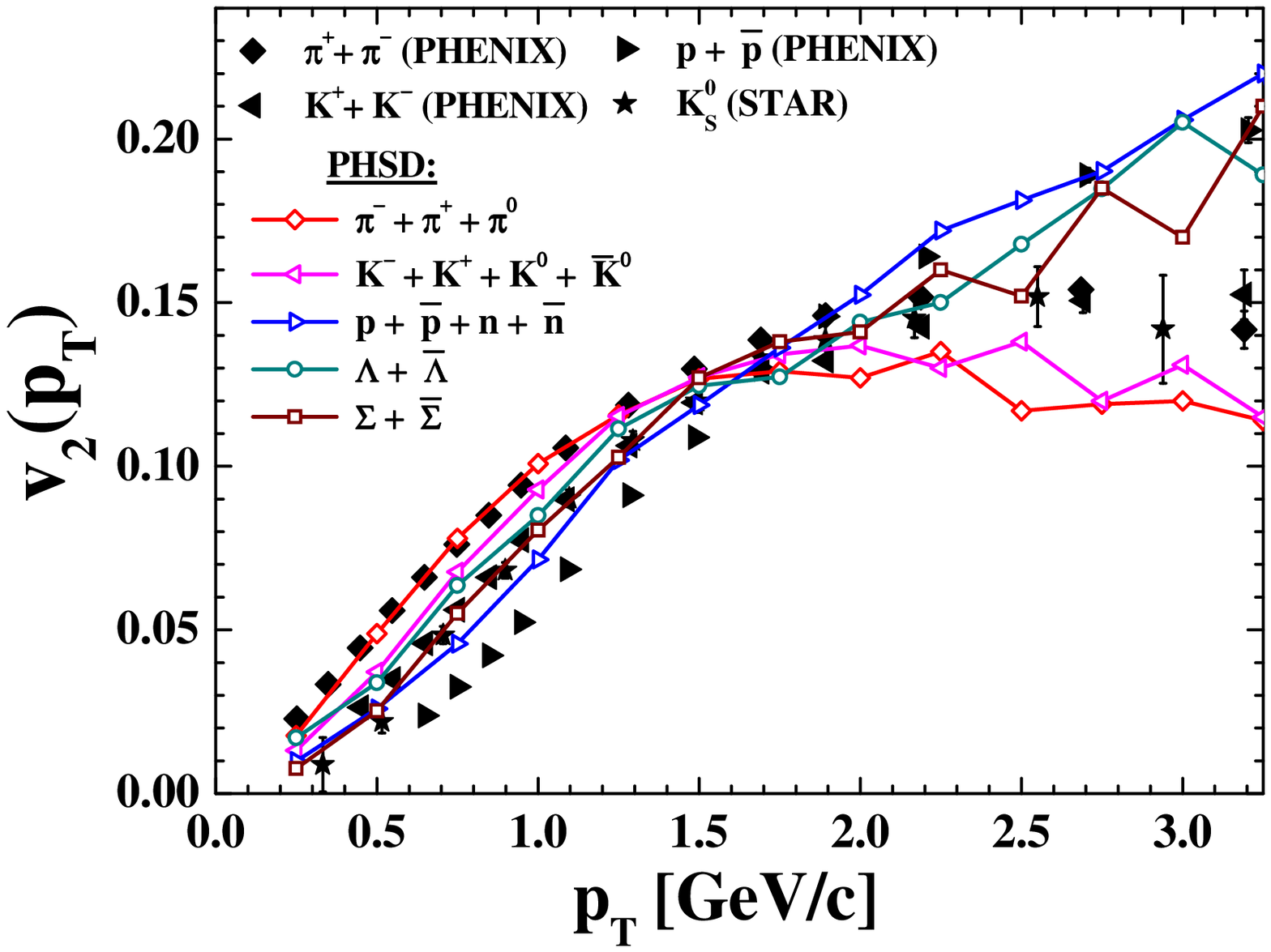}
    \includegraphics[width=0.5\textwidth]{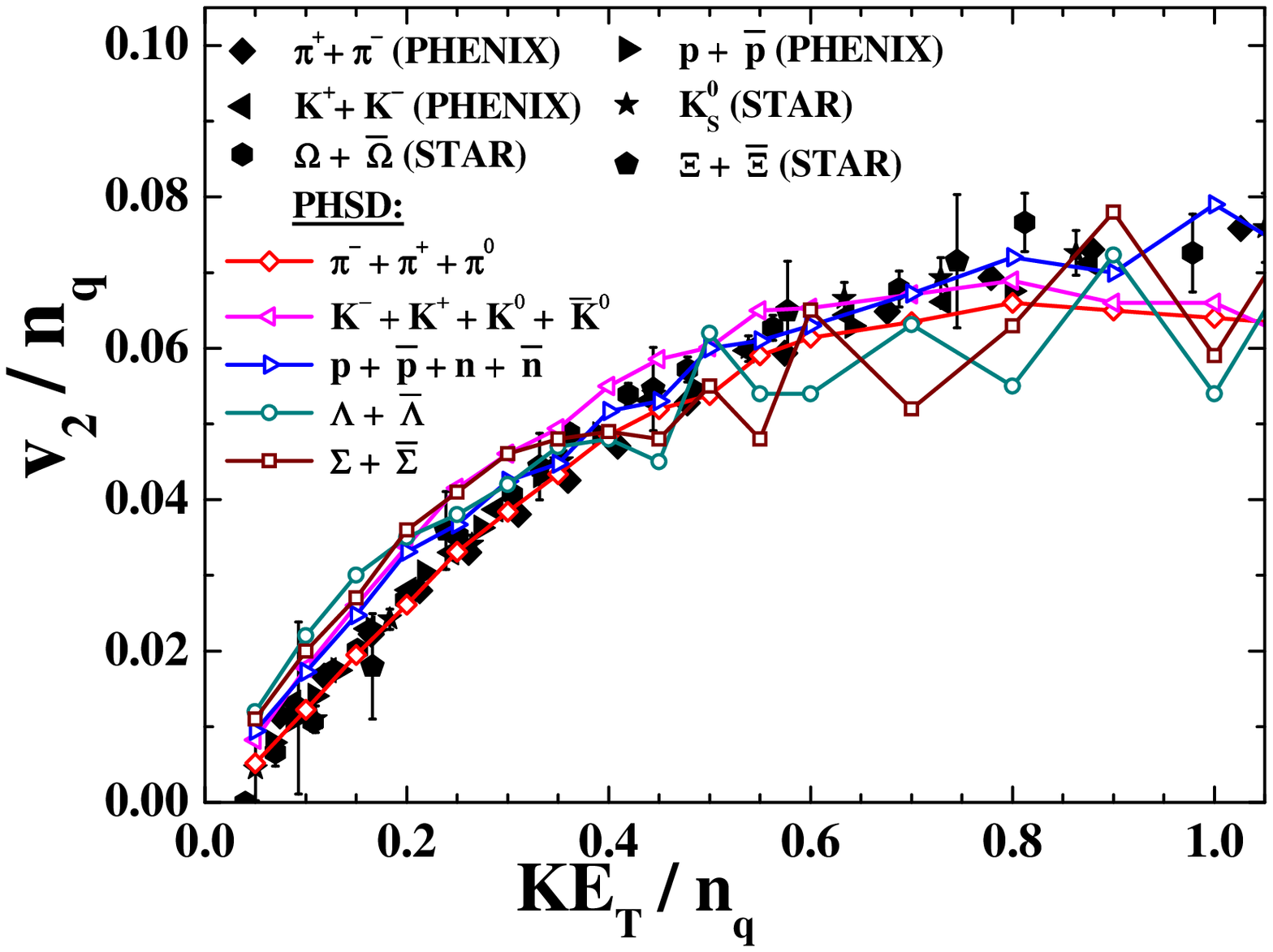}}
  \caption{ {\it Left:} The hadron elliptic flow $v_2$ for inclusive
    Au+Au collisions as a function of the transverse momentum $p_T$
    (in GeV) for different hadrons in comparison to the data from the
    STAR~\cite{STAR6} and PHENIX Collaborations~\cite{SCALING2} within
    the same rapidity cuts.  
    {\it Right:} The elliptic flow $v_2$ - scaled by the number of
    constituent quarks $n_q$ - versus the transverse kinetic energy
    (\ref{ETR}) divided by $n_q$ for different hadron species in
    comparison to the data from the STAR~\cite{STAR6} and 
    PHENIX~\cite{SCALING2} Collaborations. }
  \label{flow}
\end{figure}

Fig.~\ref{flow} (l.h.s.) shows the final hadron $v_2$ versus the
transverse momentum $p_T$ for different particle species in comparison
to the data from the STAR~\cite{STAR6} and PHENIX
Collaborations~\cite{SCALING2}. We observe a mass separation in $p_T$
as well as a separation in mesons and baryons for $p_T >$ 2 GeV
roughly in line with data. The elliptic flow of mesons is slightly
underestimated for $p_T >$ 2 GeV in PHSD which is opposite to ideal
hydrodynamics which overestimates $v_2$ at high transverse momenta. On
the other hand, the proton (and antiproton) elliptic flow is slightly
overestimated at low transverse momenta $p_T < $ 1.5 GeV.

A further test of the PHSD hadronization approach is provided by
the  'constituent quark number scaling' of the elliptic flow $v_2$
which  has been observed experimentally in central Au+Au
collisions at RHIC~\cite{STARS,SCALING2,SCALING1}. In this respect
we plot $v_2/n_q$ versus the transverse kinetic energy per
constituent parton,
\begin{equation} 
\label{ETR} KE_T = \frac{m_T-m}{n_q} \ , 
\end{equation}
with $m_T$ and $m$ denoting the transverse mass and actual hadron
mass, respectively. For mesons we have $n_q = 2$ and for
baryons/antibaryons $n_q=3$. The results for the scaled elliptic flow
are shown in Fig.~\ref{flow} (r.h.s.) in comparison to the data from
the STAR~\cite{STAR6} and PHENIX Collaborations~\cite{SCALING2} for
different hadrons and suggest an approximate scaling. For
$KE_T>0.5$~GeV there is a tendency to underestimate the experimental
measurements for $\Lambda, \Sigma, \bar{\Lambda}, \bar{\Sigma}$
baryons which we attribute to an underestimation of interaction terms
in PHSD for high momentum hadrons. In this respect we recall that the
momentum independence of the quasiparticle width $\gamma$ and mass $M$
is presently a rough approximation and has to be refined.  Due to the
limited statistics especially in the baryonic sector with increasing
$p_T$ this issue will have to be re-addressed with high statistics in
future.

%----------------------------------------------------------------------------
\section{Summary}

In this contribution we have addressed relativistic collisions of
Pb+Pb at SPS energies and Au+Au collisions at RHIC energies in the
PHSD approach which includes explicit partonic degrees of freedom as
well as dynamical local transition rates from partons to
hadrons~(\ref{trans}). The hadronization process conserves
four-momentum and all flavor currents and slightly increases the total
entropy since the 'fusion' of rather massive partons dominantly leads
to the formation of color neutral strings or resonances that decay
microcanonically to lower mass hadrons. Since this dynamical
hadronization process increases the total entropy the second law of
thermodynamics is not violated (as is the case for simple coalescence
models incorporating massless partons).

The PHSD approach has been also applied to nucleus-nucleus collisions
from 40 to 160 A$\cdot$GeV as well as for RHIC energies in order to
explore the space-time regions of 'partonic matter'~\cite{CaBra09}.
We have found that even central collisions at the top SPS energy of
$\sim$158 A$\cdot$ GeV show a large fraction of non-partonic, i.e.
hadronic or string-like matter, which can be viewed as a 'hadronic
corona'~\cite{Aichelin}.  This finding implies that neither purely
hadronic nor purely partonic 'models' can be employed to extract
physical conclusions in comparing model results with data.  On the
other hand - studying in detail Pb+Pb reactions at SPS energies in
comparison to the data~\cite{CaBra09} - it is found that the partonic
phase has only a very low impact on the longitudinal rapidity
distributions of hadrons but a sizeable influence on the
transverse-mass distribution of final kaons due to the partonic
interactions.  The most pronounced effect is seen on the production of
multi-strange antibaryons due to a slightly enhanced $s{\bar s}$ pair
production in the partonic phase from massive time-like gluon decay
and a more abundant formation of strange antibaryons in the
hadronization process. This enhanced formation of strange antibaryons
in central Pb+Pb collisions at SPS energies by hadronization supports
the early suggestion by Braun-Munzinger and Stachel~\cite{PBM,PBM2} in
the statistical hadronization model - which describes well particle
ratios from AGS to RHIC energies.

At RHIC energies the PHSD calculations show also a good reproduction
of the hadron transverse mass and rapidity spectra. Furthermore, the
elliptic flow $v_2$ is well described for Au+Au reactions at
$\sqrt{s}=200$~GeV as a function of centrality as well as of
transverse momenta up to $p_T\simeq 1.5$~GeV/$c$. Due to the local
transition rates from partons to hadrons~(\ref{trans}) the PHSD
approach also gives approximate quark number scaling of the elliptic
flow as found experimentally by the RHIC Collaborations.

\section*{Acknowledgements}
Work supported in part by the HIC for FAIR framework of the LOEWE
program and by DFG.

\end{document}